\documentclass[twocolumn]{aastex631}

\usepackage{amsmath}

\begin{document}

\title{Detection of Dipole Modulation in CMB temperature anisotropy maps\\ from WMAP and Planck using Artificial Intelligence}

\correspondingauthor{Rajib Saha}
\email{rajib@iiserb.ac.in}

\author[0000-0002-4607-5820]{Md Ishaque Khan}
\affiliation{Department of Physics, Indian Institute of Science Education and Research (IISER) Bhopal,\\
Bhopal-462066, Madhya Pradesh, India}

\author{Rajib Saha}
\affiliation{Department of Physics, Indian Institute of Science Education and Research (IISER) Bhopal,\\
Bhopal-462066, Madhya Pradesh, India}




\begin{abstract}

Breakdown of rotational invariance of the primordial power spectrum manifests in the statistical anisotropy of the observed Cosmic Microwave Background (CMB) radiation. Hemispherical power asymmetry in the CMB may be caused due to a dipolar modulation, indicating the presence of a preferred direction. Appropriately re-scaled local variance maps of the CMB temperature anisotropy data effectively encapsulate this dipolar pattern. As a first-of-its-kind method, we train Artificial Neural Networks (ANNs) with such local variances as input features to distinguish statistically isotropic CMB maps from dipole modulated ones. Our trained ANNs are able to predict components of the amplitude times the unit vector of the preferred direction for mixed sets of modulated and unmodulated maps, with goodness of fit ($R^2$) scores $>0.97$ for full sky, and $>0.96$ for partial sky coverage. On all observed foreground-cleaned CMB maps, the ANNs detect the dipolar modulation signal with overall consistent values of amplitudes and directions. This detection is significant at $97.21\%-99.38\%$ C.L. for all full sky maps, and at $98.34\%-100\%$ C.L. for all partial sky maps. Robustness of the signal holds across full and partial skies, various foreground cleaning methods, inpainting algorithms, instruments and all the different periods of observation for Planck and WMAP satellites. The significant and robust detection of the signal, in addition to the consistency of values of amplitude and directions, as found independent of any pre-existing methods, further mitigates the criticisms of look-elsewhere effects and a posteriori inferences for the preferred dipole direction in the CMB.

\end{abstract}



\section{Introduction} \label{sec:intro}
Departures from Statistical Isotropy (SI) of the Cosmic Microwave Background (CMB) temperature field may indicate limitations or errors in measurement of the CMB despite the use of highly precise instruments for observation, if not due to an actual breakdown of the rotational invariance of the primordial power spectrum. However, by means of appropriate statistical methods, systematic effects or foreground residuals may be considerably eliminated as possible causes of deviations from SI.

Several such departures from SI have been studied by authors in existing literature. These include the unusually low cosmic quadrupole \citep{Bennett_2003,10.1046/j.1365-2966.2003.07067.x, PhysRevD.69.063516}, and planarity of the cosmic octupole and the quadrupole-octupole alignment as investigated by \cite{PhysRevD.69.063516, Tegmark:2003ve,deOliveira-Costa:2003utu,Schwarz:2004gk}. The quadrupole-octupole alignment was seen to get strengthened on removal of the frequency dependent kinetic Doppler quadrupole \citep{Notari_2015}. The low multipole regime was studied by \cite{PhysRevLett.93.221301} with the help of multipole vectors and found to be consistently anomalous with respect to multipole aligments. Further, \cite{PhysRevLett.95.071301} showed that a mysterious correlation exists between azimuthal phases of the third and fifth multipole moments. A significant power asymmetry between the two hemispheres of the CMB was found by \cite{Eriksen_2004, 10.1111/j.1365-2966.2004.08229.x,Eriksen_2007} and further corroborated by \cite{Bernui_2014}.

A power excess for odd multipoles was studied in the work of \cite{PhysRevD.72.101302}. This parity asymmetry in the CMB angular power spectrum (APS) was confirmed by \cite{Kim_2010, PhysRevD.82.063002} and the anomaly was seen to disappear without the contribution of the first six low multipoles \citep{10.1111/j.1365-2966.2011.19981.x}. Using symmetry-based methods of power and directional entropy statistics, \cite{10.1111/j.1365-2966.2008.12960.x,  10.1111/j.1365-2966.2009.14728.x} showed that the departures from SI extend to higher multipoles as well. For scales above $60^\circ$, nearly negligible correlation was seen by \cite{PhysRevD.75.023507, 10.1111/j.1365-2966.2009.15270.x,  10.1093/mnras/stv1143} with various CMB data releases. \cite{kim_lack_2011} showed that the occurrence of parity asymmetry in the APS is equivalent to this deficit of large angle correlation.  Further, on the basis of behaviours of level clustering and repulsion for uncorrelated and correlated values, \cite{Ishaque_Khan_2022} showed that only the level correlations between even multipoles is anomalously low. \cite{PhysRevD.89.023010} studied a directional dependence of the parity asymmetry and suggested a common origin of the low multipole anomalies.

Additionally \cite{Larson_2004} showed that the mean values of hot and cold spots of the CMB are unexpectedly low, while \cite{10.1111/j.1365-2966.2008.13149.x} found that the variance of the CMB temperature anisotropy field is also anomalously low. The low CMB variance anomaly was seen to vanish when the quadrupole and octupole were excluded from the CMB maps under investigation \citep{10.1111/j.1365-2966.2010.18067.x}. Using novel statistics to measure the strength and shape of distribution of CMB local extrema, \cite{Khan_2022} found a strikingly weak non-uniformity in the distribution of hot and cold spots on the CMB, which is due to the low CMB temperature variance and anomalous contributions of the quadrupole and octupole.

It is important to investigate any CMB anomaly from as many perspectives as possible to assess its significance and role in cosmological parameter estimation. For example, the direction associated with CMB parity asymmetry aligns at about $45^\circ$ from a best-fit dipole form for various cosmological parameters \citep{PhysRevD.105.083508}. Besides, a directional variation of the cosmological parameters on the CMB sky was found to be significantly anisotropic and this finding is corrrelated with the preferred direction for the hemispherical power asymmetry anomaly \citep{10.1093/mnras/stab1193}. Since these works report a correlation between these departures from SI of the CMB and the anisotropic directional dependence of cosmological parameters, hence it becomes difficult to disregard the violations of SI as mere statistical fluctuations \citep{https://doi.org/10.48550/arxiv.2207.05765}.

These departures from SI were found to be robust against masking of the CMB sky, instruments used for observation, foreground cleaning methods, periods of observation, bands of frequencies at which the CMB is observed, and the like. Further, checks of robustness help reduce the possibility that the significant results can be attributed to look-elsewhere effects. Many such independently conducted findings of deviations from SI also weaken the inference that the consequent signal detection could have happened solely due to the nature of estimators which were designed by hand `a posteriori' \citep{Bennett_2013, 2014JCAP...08..006R} to focus on some unusual features.

However, despite the high statistical significance of most of such departures from SI, they are ascertained to be fairly within the underlying probability distribution given by the $\Lambda CDM$ model. Thus we can have either of two possible conclusions: (a) we may say that we happen to inhabit a rare realisation of the universe given by the $\Lambda CDM$ standard model, or (b) we inhabit a reasonably probable realisation of a different model. The latter case then warrants contemplation of new physics beyond the Standard Model of Cosmology.

One of such departures from SI which has been robustly observed, is the hemispherical power asymmetry \citep{Eriksen_2004, Eriksen_2007}. It was hypothesised to be engendered by the addition of a dipolar modulation to otherwise statistically isotropic CMB temperature anisotropy fluctuations $T_0(\hat{n})$. Thus, the net temperature anisotropies in this scenario are
\begin{eqnarray}\label{tmod}
    T(\hat{n})&=&T_0(\hat{n})\left(1+A\hat{\lambda}\cdot\hat{n}\right) \enspace ,
\end{eqnarray} where, the amplitude of modulation is denoted by $A$, and the preferred direction is given by the unit vector $\hat{\lambda}$. In harmonic space, the temperature fluctuations $T_0(\hat{n})$ are decomposed as:
\begin{eqnarray}
    T_0(\hat{n})&=&\sum_{l=0}^\infty\sum_{m=-\ell}^{\ell} a_{\ell m} Y_{\ell m} (\hat{n}) \enspace .
\end{eqnarray}These $T_0(\hat{n})$ are expected to be Gaussian random and generated from a rotationally invariant primordial power spectrum. Hence there are no preferred directions in the standard model that may couple modes of these temperature fluctuations in harmonic space. This notion of SI is encapsulated in the relation
\begin{eqnarray}\label{SI}
    \langle a_{\ell m}  a _{\ell' m'}^*\rangle &= & C_\ell \delta_{\ell \ell'} \delta_{m m'} \enspace .
\end{eqnarray}
Thus the spherical harmonic coefficients $a_{\ell  m}$ are uncorrelated between different multipoles. However, if the dipole modulated $T(\hat n)$ are similarly decomposed in spherical harmonics, the corresponding $a_{\ell m}$ will contain correlations between multipoles $\ell$ and $\ell+1$ \citep{PhysRevD.91.023515}, indicating a violation of SI.

As an entirely novel approach towards understanding the possible presence of a dipolar modulation in CMB temperature anisotropy data, we employ Artificial Neural Networks (ANNs). ANNs are computer based analogs of networks of biological neurons, and constitute an important machinery with decision making and parameter estimation capabilities, that falls under the umbrella of Artificial Intelligence (AI). We use deep learning techniques to train the ANNs on a mixed set containing equal numbers of simulated SI obeying (unmodulated) and SI violating (dipole modulated) CMB maps, which is inclusive of a large number of possibilities of the presence or absence of the signal. Thus our trained ANNs can make a self-guided and robust estimation of the presence of the signal of dipolar modulation, quantified with the value of the amplitude. The rationale behind using the amplitude for this purpose is that CMB maps that obey SI will have zero amplitude for such modulation, whereas those that contain the modulation will have non-zero values of the amplitude. As a realistic approach, we design an ANN with partial sky coverage in addition to one that works for full sky coverage, since we may not always have completely reliable full sky observations. Besides, we are able to compute the directions of the modulation with the help of the trained ANNs. Thus our method serves as an independent investigation to establish or reject the existence of the dipolar modulation signal as seen in existing literature. 

Previously, statistics or estimators have been devised to ascertain the amplitude and direction of a possible dipolar modulation in the CMB. Estimators can be constructed in pixel or harmonic space, as per the requirements of the studies that undertake the same. For example, since the amplitude of the modulation has been shown to be dependent on the scale \citep{Hoftuft_2009} and hence the multipole range under consideration, studying estimators in multipole space helps estimate this scale dependence \citep{Marcos-Caballero_2019, Pranati_K_Rath_2013}. Whereas, an analysis in pixel space can be immensely useful so as to avoid subtle biases introduced due to masking of the sky that causes extraneous couplings in multipole space \citep{Hivon_2002}, or those caused due to inpainting of partial sky maps \citep{inpaint_Starck}. In this work, we train ANNs with normalised or re-scaled local variance maps \citep{Akrami_2014} in pixel space, which serve as important input features containing direct information of the amplitude and directions of the dipolar modulation in the form of scalar products. This method helps us eschew the complex task of constructing statistics for detection of the signal. The ANNs are designed to work on scales of observation corresponding to the range of multipoles $\in[2,256]$. We defer a study of the scale dependence of $A$ to future work.

The implementation of ANNs for detecting previously studied features in the CMB could revolutionise perspectives towards understanding CMB anomalies as opposed to classical fitting or regression methods and traditional frequentist approaches. ANN architectures can `learn' signal detection capabilities by being introduced to a training set of samples. Once trained, the ANN can then be fed observed foreground-cleaned CMB data to predict a possible signal in the same.

A comprehensive review of the preliminary use of ANNs in Astronomy and Astrophysics can be found in the article by \cite{MILLER1993141} with regard to telescope optics, object classification and filtering of detector events. Further \cite{Tagliaferri2003-vd, Wang2018, https://doi.org/10.48550/arxiv.2212.01493, CHEN2020347} describe the growth of ANN based algorithms to perform time series analysis, detection of noise, and data mining in addition to classification and identification of astrophysical objects such as new stars, galaxies or even dark matter.

In Cosmology, use of ANNs has ushered in a new era of numerical frameworks to ease computations and analyses. They were used by \cite{Liu_2017} for generating dynamics of inflationary trajectories in a multi-field scenario. \cite{Dialektopoulos_2022} used ANNs to reconstruct late-time expansion and large scale structure (LSS) cosmological parameters. \cite{Wang_2020} used them to estimate quantities such as the Hubble parameter and luminosity distance as a function of redshift of Type Ia supernovae.  \cite{Gomez-Vargas_2021} modelled ANNs with Bayesian inference to calculate the likelihood function and reduce computation time for cosmological parameter estimation.  \cite{Escamilla-Rivera_2020} provided a combined Bayesian and Recurrent neural network approach to ascertain confidence regions for parameters from dark energy models. Besides, ANNs can be designed for estimation of parameters using the $21$ $cm$ signal from the epoch of reionization \citep{10.1093/mnras/stx734, 10.1093/mnras/stab180}. A general overview of ANNs and their applications in analysis of cosmological data can be found in the article by \cite{universe8020120}.

Recent applications of ANNs specific to CMB data analysis can be found in the following works. \cite{Petroff_2020} implemented an appreciable full-sky foreground cleaning of the observed CMB, while \cite{Wang_2022} were able to recover CMB signals from foreground contaminated maps using Convolutional neural networks (CNNs). \cite{10.1093/mnras/stab2753} applied ANNs to successfully recover full sky CMB temperature APS from a low resolution masked or partial sky CMB map, while \cite{https://doi.org/10.48550/arxiv.2203.14060} designed ANNs for such estimation of full sky CMB temperature power spectrum with higher resolution partial sky CMB maps. Using CNNs \cite{https://doi.org/10.48550/arxiv.2211.09112} reconstructed the full sky power spectra of CMB $E$ and $B$ modes for such high resolution CMB maps, while minimising the leakage between the two modes. \cite{10.1111/j.1365-2966.2011.20288.x} implemented Bayesian inference algorithms to make ANNs learn the likelihood function and estimate cosmological parameters from CMB data. \cite{10.1093/mnras/staa1469, https://doi.org/10.48550/arxiv.2011.14276} trained ANNs to mimic mixing of Markov chains (MCs) and parameterization of Monte Carlo MC proposals. \cite{10.1093/mnras/stac064} developed ANN based estimators to compute the matter and CMB power spectra as a replacement of Boltzmann codes suited for both LSS and CMB surveys.

We have organised our paper as follows. In Section \ref{formalism} we present the underlying formalism behind normalised local variance maps which can be directly used as input features for training a neural network. In Section \ref{ANN_info}, we briefly describe the internal structures of ANNs and the algorithms with which they function as trainable artificial analogs of biological neural networks. We elucidate our procedure for obtaining mixed sets of unmodulated and modulated CMB maps, and using them for training the ANNs in Section \ref{methodology}. Following this, we discuss the specific structure of our ANNs and regularization methods used to train the same for both full and partial sky maps in Section \ref{train_process}. The analysis of test sets and observed foreground-cleaned CMB maps are presented in Section \ref{analysis}, after application of our trained ANNs to those. In Section \ref{conc}, we summarise our work, and enumerate the key findings of the paper.

\section{Formalism}\label{formalism}

For the CMB temperature anisotropy field defined on the 2-sphere of observation, we can compute its variances inside different local regions of the sphere. We consider these regions to be discs of equal area spanning the 2-sphere. In this Section, we present the formalism of how such local variances, after appropriate re-scaling are equivalent to the amplitude times a scalar product of the constant unit vector of modulation and the mean direction of the local disc, as first utilised in the work of \cite{Akrami_2014}.

In HEALPix \citep{2005ApJ...622..759G} notation, the parameter $n_{side}$ characterises the pixel resolution of a CMB map. For convenience, we denote the $n_{side}$ of a high resolution CMB map by $n_{h}$ and that of a lower resolution CMB map by $n_l$. We consider a disc of radius $r_h$, on the map at resolution $n_h$. We then calculate the local mean and variance of temperature fluctuations within the disc. Thus the mean of modulated disc temperature fluctuations from equation \eqref{tmod} is,
\begin{eqnarray}\label{meanT}
    \langle T\rangle_d &=& \langle T_0\rangle_d+\langle A T_0\hat{\lambda}\cdot\hat{n}\rangle_d \enspace ,
\end{eqnarray}
where $\langle\rangle_d$ denotes expectation value over the disc. Let us consider the second term in the equation above where $A$ is a constant. The statistically isotropic Gaussian random fluctuations $T_0$ are approximately independent of and uncorrelated with the variations of $\hat{\lambda}\cdot\hat{n}$, if the disc is sufficiently small enough so that the $\hat{\lambda}\cdot \hat{n}$ term is slowly varying. Hence, the expression \eqref{meanT} reads:
\begin{eqnarray}
    \langle T\rangle_d &=& \langle T_0\rangle_d+A\langle T_0\rangle_d\langle\hat{\lambda}\cdot\hat{n}\rangle_d \enspace .
\end{eqnarray}
The local variance within the disc is
\begin{eqnarray}
    \sigma^2_d &=& \langle\left(T-\langle T\rangle_d\right)^2\rangle_d \enspace .
\end{eqnarray}
Expanding out, the expression for the disc variance becomes
\begin{eqnarray}
    \sigma^2_d &=& \langle\left[\left(T_0-\langle T_0\rangle_d\right)+A\left(T_0\hat{\lambda}\cdot\hat{n}-\langle T_0\rangle_d\langle \hat{\lambda}\cdot\hat{n}\rangle_d\right)\right]^2\rangle_d \nonumber\\
    &=& \langle\left(T_0-\langle T_0\rangle_d\right)^2\rangle_d \nonumber \\
    && + 2A\langle \left(T_0-\langle T_0\rangle_d\right)\times\left(T_0\hat{\lambda}\cdot\hat{n}-\langle T_0\rangle_d \langle\hat{\lambda}\cdot\hat{n}\rangle_d\right)\rangle_d \nonumber \\
    && + \mathcal{O}(A^2) \nonumber \\
    &=& \sigma^2_{0d}+ 2A\Big [\langle T_0^2 \hat{\lambda}\cdot\hat{n}\rangle_d-\langle T_0 \langle T_0\rangle_d\langle\hat{\lambda}\cdot\hat{n}\rangle_d\rangle_d \nonumber \\
    && -\langle T_0\langle T_0\rangle_d \hat{\lambda}\cdot\hat{n}\rangle_d + \langle T_0\rangle_d^2\langle\hat{\lambda}\cdot\hat{n}\rangle_d\Big ] +\mathcal{O}(A^2)\nonumber \\
    &=& \sigma^2_{0d} + 2A\left[ \langle T_0^2\rangle_d-\langle T_0\rangle_d^2\right]\times\langle \hat{\lambda}\cdot\hat{n}\rangle_d +\mathcal{O}(A^2) \enspace ,
\end{eqnarray}where $\sigma_{0d}^2$ stands for the disc variance in the absence of a modulation. We can replace the term $\langle \hat{\lambda}\cdot\hat{n}\rangle_d$ with $\hat{\lambda}\cdot\langle\hat{n}\rangle_d$, since $\hat{\lambda}$ for any particular CMB map is constant. Further, as the average of position vectors $\hat{n}$ is over a disc, $\langle\hat{n}\rangle_d=\hat{N}$, where $\hat{N}$ is the centre of the disc. Hence,
\begin{eqnarray}
    \frac{\sigma^2_d-\sigma^2_{0d}}{\sigma^2_{0d}} &\simeq & 2A \hat{\lambda}\cdot \hat{N} \enspace .
\end{eqnarray}
However, evaluating $\sigma_{0d}^2$ from a single Gaussian random realisation may give a biased estimate. Instead, we will compute $\langle \sigma^2_{0d}\rangle_{e}$ and use that in our expression above. This expectation $\langle \rangle_{e}$ is over an ensemble of statistically isotropic realisations. Thus finally we arrive at the following normalised local variance (NLV),
\begin{eqnarray}
    \boxed{\frac{\sigma^2_d-\langle\sigma^2_{0d}\rangle_{e}}{\langle\sigma^2_{0d}\rangle_{e}} \simeq  2A \hat{\lambda}\cdot\hat{N}} \enspace ,
\end{eqnarray}which shall be used in all analyses hereafter. 

Several discs on the $n_{h}$ map are considered and their NLVs computed. These NLV values are then assigned to corresponding pixels of another map at a lower resolution $n_l$. Thus to construct an NLV map, the total number of discs to be considered $=12\times n_{l}^2$. The centre of any particular disc on the $n_{h}$ map is taken to be the same as the position vector of the pixel of the $n_{l}$ resolution map to which the NLV of that disc is assigned. With this information, we can calculate the approximate number of pixels ($n_{pd}$) of the $n_{h}$ map inside each disc of given radius $r_h$ (in degrees). This is expressed as
\begin{eqnarray}\label{npxd}
    n_{pd}&=&\frac{\text{Area of the disc}}{\text{Area of a pixel in $n_{h}$}} \nonumber \\
    &=& \pi\left(\frac{r_h\times \pi}{180}\right)^2 \Big / \frac{4\pi}{12\times n_{h}^2} \enspace , \nonumber
\end{eqnarray}
\begin{eqnarray}
    \boxed{n_{pd} = 3\times \left(\frac{n_{h}\times r_h\times\pi}{180}\right)^2} \enspace .
\end{eqnarray}We can calculate the approximate number of pixels in possibly overlapping regions as follows. For $n_{pl}=12\times n_{l}^2$ discs, the total number of pixels taken is $n_{pt}=n_{pl}\times n_{pd}$. Thus if $n_{pt} > n_{ph}$, then there are $n_{pt}-n_{ph}$ number of pixels which are present in overlapping regions of discs, and vice versa. Here, $n_{ph}=12\times n_{h}^2$.

\section{How does an ANN work?}\label{ANN_info}

\begin{figure*}
    \centering
    \includegraphics[width=0.8\textwidth]{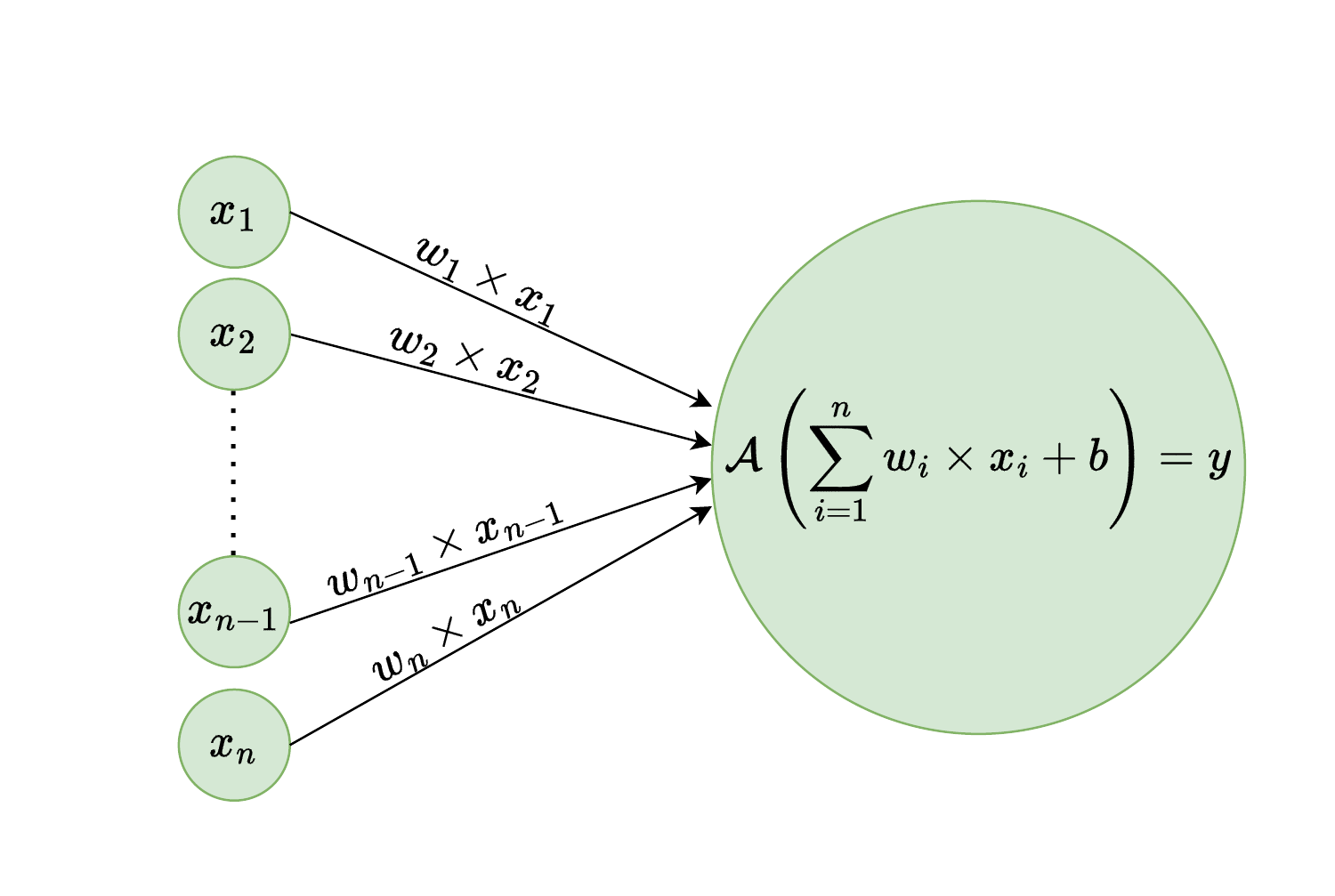}
    \caption{An artificial neuron is the building block of an ANN. Its structure comprises inputs $x_i$'s which are weighted with $w_i$'s, summed over and added to a bias $b$, the resultant of which is acted on by an activation function $\mathcal{A}$ to give the output $y$.}
    \label{neuron1}
\end{figure*}
\begin{figure*}
    \centering
    \includegraphics[width=0.8\textwidth]{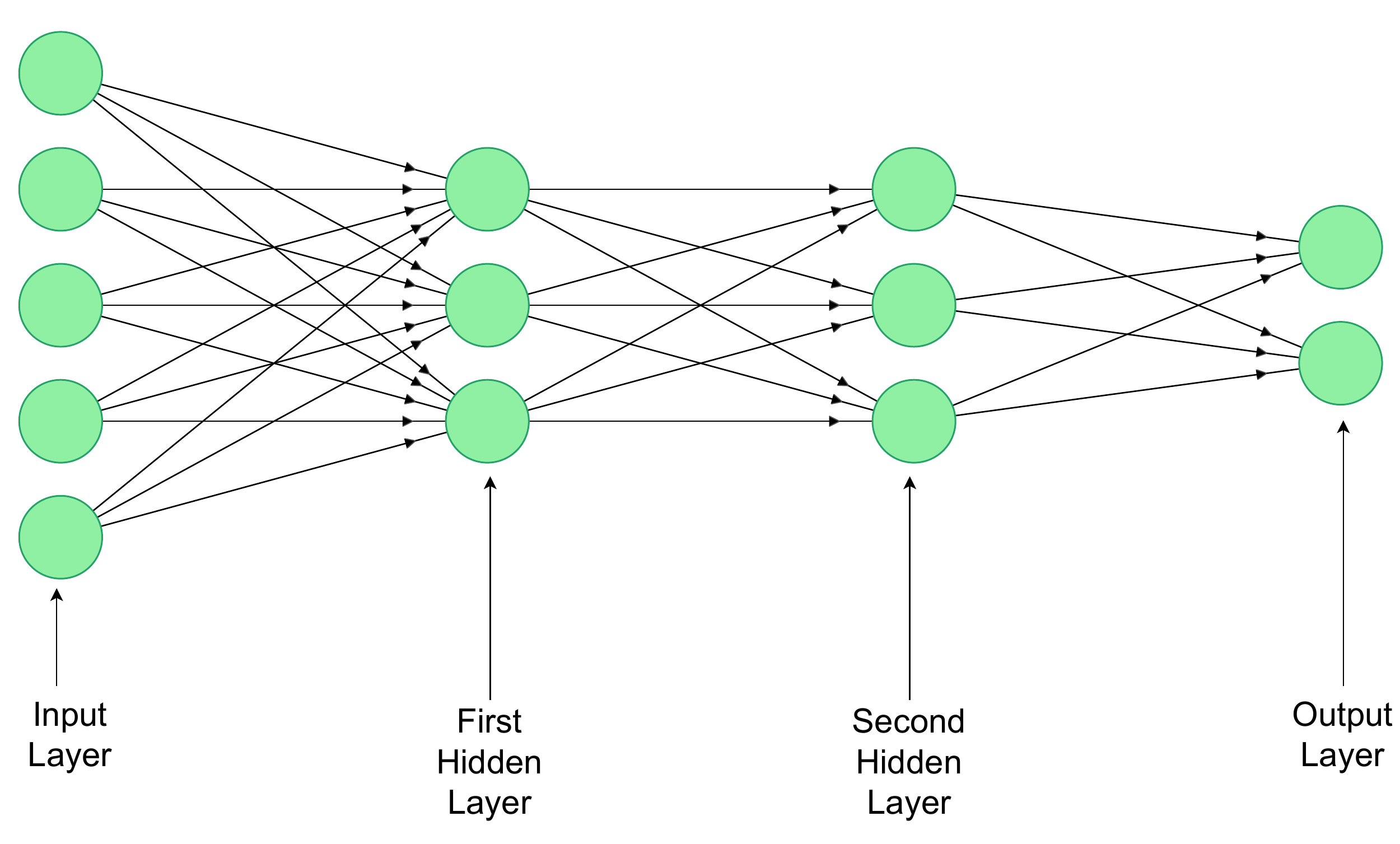}
    \caption{An example of an ANN architecture, containing $5$ inputs, two hidden layers with $3$ nodes each, and $2$ outputs. Each layer after the input layer has nodes which are densely connected to those of the preceding layer.}
    \label{egANN}
\end{figure*}

An artificial neuron is the building block of an ANN, which is inspired from the concept of the biological neuron \citep{mcculloch_pitts_1943, Rosenblatt1958ThePA}. It combines a weighted sum of several inputs, adds a bias to that sum to give a preliminary output. Since this preliminary output is a linear mapping, no matter how many neurons are interconnected to form a network, the mapping from initial inputs to the last output can always be described as a linear mapping \citep{M2016-jo}. In reality, the complex relations between inputs and expected outputs may never be reducible to a linear mapping. Hence using activation functions \citep{https://doi.org/10.48550/arxiv.2109.14545} to introduce non-linearity becomes pertinent. Thus a modern day artificial neuron can be represented by the example in Figure \ref{neuron1}, which shows how inputs $x_i$ are weighted by $w_i$ and summed together along with an offset or bias $b$. An activation function $\mathcal{A}$ acts on this sum to give the subsequent output $y$.

The initial inputs $x$ form a layer of nodes called the `input layer'. Several such outputs $y$ can be formed from these initial inputs, using different weights and biases. All these $y$ are then a new set of nodes that constitute the first `hidden layer'. The second hidden layer can be formed by considering $y$'s from the first hidden layer as inputs, and so on. A small ANN may have one or two hidden layers, whereas a larger ANN could comprise several such hidden layers. The last layer of any ANN consists of the final outputs and is called the `output layer'. The number of nodes in this layer is equal to number of expected outputs that the ANN is being trained for. Due to the presence of hidden layers that are densely connected to their preceding layers, the method of training such ANNs is referred to as `deep learning' \citep{10.3389/frai.2020.00004}.

The activation function used to obtain the outputs usually differs from the choices in other hidden layers. For example, while the $ReLU$ or $LeakyReLU$ activation functions can be used for hidden layers, for the output layer, the respective activation used could be a $sigmoid$ or $softmax$ for binary or multi-class classification problems, or $linear$ for regression problems \citep{https://doi.org/10.48550/arxiv.2101.09957}. To illustrate the arrangement of a neural network, we show an example ANN in Figure \ref{egANN}. For all layers apart from the input layer, each node is fully connected to all the nodes in the previous layer.

Beginning with the input layer as the $0^{th}$  layer, we can successively number the other layers. Consider the $i^{th}$ node in the $(l-1)^{th}$ layer, which is connected to the $j^{th}$ node in layer $l$. The associated weight and bias will be $w_{ij}^{(l)}$ and $b_{j}^{(l)}$. Thus the value taken by the $j^{th}$ node in layer $l$ is:
\begin{eqnarray}
    y_j^{(l)} &=& \mathcal{A}\left(\sum_i w_{ij}^{(l)}\times y_{i}^{(l-1)}+b_j^{(l)}\right).
\end{eqnarray}If $l=1$, then $y_j^{(l)}$ represents a node in the first hidden layer and $y_i^{(l-1)}=x_i$ corresponds to that of the input layer. The 2D weight matrix between layers $(l-1)$ and $l$ is given by $[W^{(l)}]_{ij}=w_{ij}^{(l)}$, while the bias column vector is $[B^{(l)}]_{j}=b_j^{(l)}$. To begin with, the weights and biases for the ANN can be chosen randomly.

In the process of assigning values to nodes in subsequent layers, a forward propagation in the ANN is achieved. To verify if the final outputs are as expected, a loss function is computed for the outputs generated \citep{Goodfellow-et-al-2016}. Since in this paper, we are dealing with a regression problem, we will consider the loss function as the $mse$ or mean squared error, given as
\begin{eqnarray}
    mse &=& \frac{1}{N}\times \sum (y_{true}-y_{pred})^2 \enspace ,
\end{eqnarray} where the summation is over a total number of output values $N$. Thus $mse$ is the average of squared differences between the predicted outputs ($y_{pred}$) from the ANN and the true values ($y_{true}$). To train the ANN effectively, we must perform back-propagation \citep{rumelhart_hinton_williams}, in which the weights and biases associated with the different layers are updated iteratively so as to minimize the $mse$ loss function. The rate or step-size for updating in this process is known as the learning rate $\sigma$. Thus the basic relations which describe the process of back-propagation for a loss function $H$ are,
\begin{eqnarray}
    W^{(l)} &\to & W^{(l)} - \sigma\times \nabla_{W^{(l)}} H \enspace , \nonumber \\
     B^{(l)} &\to & B^{(l)} - \sigma\times \nabla_{B^{(l)}} H \enspace .
\end{eqnarray}These relations correspond to the algorithm of gradient descent. However, such an algorithm when applied to the whole data-set can be computationally expensive. Thus the data set is divided into several batches \citep{inproceedings_batch_size} randomly, and the above algorithm is applied. A batch represents the number of samples from the data-set which are used during a part of an iteration for updating the parameters of weights and biases. A complete iteration during which the whole data-set is made to undergo the algorithm for optimization is called an epoch. Due to the random subdivision of the training set into batches, some stochasticity is introduced into the loss function. For our problem of regression of the amplitude and directions of any possible dipolar modulation, we have considered the $Adam$ (adaptive moment estimation) optimizer \citep{https://doi.org/10.48550/arxiv.1412.6980}, which incorporates adaptive estimates of the gradients and their squares. In this method, the parameters (weights and biases) are updated as follows.
\begin{enumerate}
    \item A step-size or learning rate $\sigma$ is specified.
    \item Exponential decay rates for the moment estimates are $\gamma_1$, $\gamma_2$ $\in [0,1)$.
    \item Stochastic loss function $H(\delta)$ is given.
    \item Parameters $\delta$ are initialised randomly.
    \item The first moment vector $m_0$, second moment vector $v_0$ and time-step $t$ are initialised to zero.
    \item The time step is updated as $t \to t + 1$ .
    \item Gradient $g_t = \nabla_\delta H_t(\delta_{t-1})$ .
    \item $m_t = \gamma_1\times m_{t-1} + (1 -\gamma_1)\times g_t$ .
    \item $v_t = \gamma_2 \times v_{t-1} + (1 -\gamma_2) \times g_t^2$ .
    \item Bias correction for first moment estimate: $\hat {m_t} = m_t /(1 -\gamma_1^t)$ .
    \item Bias correction for second moment estimate: $\hat{v_t}=v_t/(1 -\gamma_2^t)$ .
    \item Updating of parameters: $\delta_t =\delta_{t-1} -\sigma\times \hat{m_t}/(\sqrt{\hat{v_t}} + \epsilon)$ .
    \item Steps $6$\textendash$12$ are repeated until $\delta_t$ converges.
    \item Resulting parameters are $\delta_t$ .
\end{enumerate}
Here, values of $\gamma_1=0.9$, $\gamma_2=0.999$, and $\epsilon=10^{-7}$. Superscripts $t$ in $\gamma_1^t, \gamma_2^t$, denote that those are raised to the power of $t$. The algorithm described above for the $Adam$ optimizer is run for each of the batches within an epoch so that the ANN trains with the entire data set during one epoch itself. Several epochs may be required for the ANN to become fully trained. 

At the end of an epoch, the ANN evaluates the final loss function from the set on which it is being trained. To infer whether or not an ANN is fully trained, i.e, if it is able to generalise its knowledge to sets on which it has not been trained, another data-set for validation is simultaneously considered at every epoch. The ANN acts on this set and generates the corresponding loss value. Depending on the nature of the loss function, the optimization of the same may either correspond to that of minimisation or maximisation. We consider the case of $mse$ as the loss function, which must be minimized. Thus, over a considerable number of epochs, if the training loss does not appear to minimize, then the ANN is said to be `under-fitting' and usually a more complex network architecture can help resolve the issue. On the other hand, if the training loss adequately reaches its minimum, while that of validation does not, then the ANN is said to be `over-fitting'. This can be seen from the loss curves, where the validation loss curve is always above the training loss, whereas the training loss has converged to an appropriate minimum. The condition of over-fitting indicates that the weights and biases in the ANN are very well suited for the training set, but those are not optimal for the ANN to make appropriate predictions for new data sets that it has not `seen' before.

Over-fitting can be resolved using regularization methods \citep{Ying_2019} such as those of the $L1$ or $L2$ penalty or with the help of a `dropout'. In designing our ANNs for estimation of dipolar modulation parameters, we have used kernel regularizers with $L1$ and $L2$ penalties in addition to a dropout. Kernel regularizers penalise the loss function of the training set by adding to it a strength factor ($\mathcal{S}$) times the penalty $\mathcal{P}$, which is formed from entries in the weight matrices. In case of the $L1$ kernel regularizer, $\mathcal{P}$ is the sum of absolute values of the weights, whereas for the $L2$ regularizer, it is the sum of the squares of the weights. On the other hand, if a dropout \citep{https://doi.org/10.48550/arxiv.1207.0580} is applied before a layer, a fraction of the inputs from the previous layer are randomly dropped out, i.e, set to zero. This fraction is known as the rate of dropout and its value lies $\in [0,1]$.

\section{Methodology}\label{methodology}

\begin{figure*}
    \centering
    \includegraphics[width=0.8\textwidth, keepaspectratio]{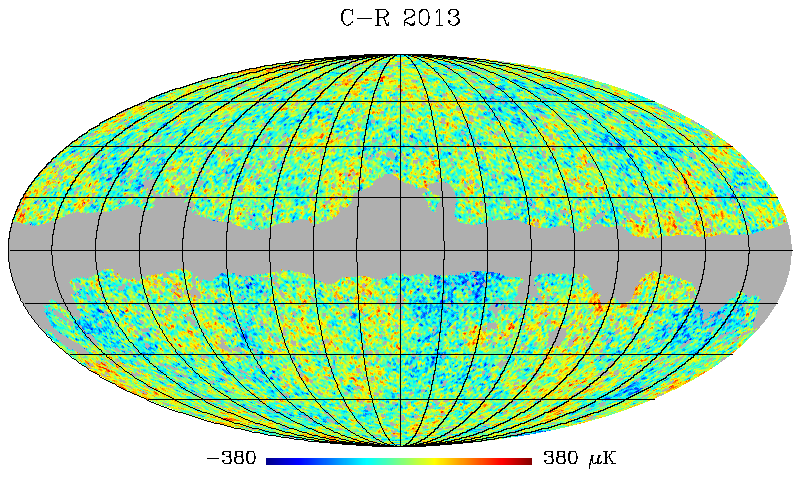}
    \caption{As an example of partial sky coverage, we present the Planck 2013 Commander-Ruler (C-R) map at resolution $n_h=128$ after application of the Planck 2013 $U73$ mask. The masked regions are shown in grey colour. The CMB fluctuations are shown in thermodynamic temperature units of $\mu K$.}
    \label{masked_map}
\end{figure*}

We have considered a mixed set of $5\times 10^4$ randomly generated CMB realisations of maps at $n_h=128$. Half of these are statistically isotropic, and the other half are dipole modulated versions of the same.

The statistically isotropic CMB temperature maps at $n_{h}=128$ are obtained by choosing the spherical harmonic coefficients $a_{\ell m}$ as Gaussian random variables with zero mean and variance given by the theoretical CMB temperature APS best fit to Planck 2018 data. Thus, we generate $2.5\times 10^4$ SI obeying maps with different seed values.

In order to obtain $2.5\times 10^4$ dipole modulated counterparts of the SI obeying maps at $n_h=128$, we utilise equation \eqref{tmod}. We pick $A$ from a uniform random distribution in accordance with the order of magnitude of reported values from observed foreground-cleaned data. We have further considered a wide range given by $A\in[0.03,0.15]$ so as to sufficiently accommodate a large number of possible values of amplitude. This further helps minimize the epistemic uncertainty \citep{Hullermeier2021} of the ANN. 

The direction of the dipole for modulation ($\hat{\lambda}$) is chosen in the following manner. For three different seed values, we generate three random numbers from a normal distribution with a mean of zero and standard deviation equal to one. The numbers are chosen such that the sum of their squares are non-zero. They are then normalised by the square root of the sum of their squares. Thus the three resulting numbers form components of the randomly chosen unit vector $\hat{\lambda}$, which gives the preferred direction of dipolar modulation for a particular realisation. The rationale behind choosing the components as random normal numbers is to take into account all possible directions on the sphere \citep{10.1145/377939.377946}, since other choices such as those of random uniform numbers restrict the randomness in directionality of the modulation.

The NLV maps at $n_{l}=16$ are constructed using discs of radius $=6^{\circ}$. Thus inside each disc, the approximate number of pixels at $n_{h}=128$ over which the local variances are computed is approximately $540$, according to equation \eqref{npxd}, for which it can be shown that there are about $1459237$ pixels in overlapping regions of discs.

The manually adopted choice of $r_h=6^{\circ}$ is an optimal one due to the following reasons. Local variance estimates over smaller disks will have relatively higher contributions from Monte Carlo noise due  to lower number of  pixels contained by them, which must be avoided. Besides, very small radii such as $r_h\lesssim 4^{\circ}$ are subject to non-negligible contributions from the Doppler dipole \citep{10.1093/mnras/stu2408}. However, choosing large radii weakens the assumption of a slow variation of $\hat{\lambda}\cdot\hat{n}$ inside the disc (Section \ref{formalism}), and can cause results to concur with statistically isotropic maps \citep{Akrami_2014} for very large $r_h$. Hence, we choose $r_h=6^\circ$ which is sufficiently small, and reasonably free from contributions of the Doppler dipole and Monte Carlo noise.

In order to construct the NLV maps, we require a mean variance map containing $\langle\sigma_{0d}^2\rangle_{e}$ values. This mean variance map is obtained using an ensemble of $1 \times 10^5$ local variance maps at $n_{l}=16$, which were extracted from the same number of corresponding SI obeying realisations of maps at $n_{h}=128$. Of the total mixed set of $5\times 10^4$ NLV maps, $2\times 10^4$ maps are used for training the ANNs, $10^4$ are used for validation, and the remaining $2\times 10^4$ are used for testing the trained ANN.

We consider two cases of sky coverage, i.e, full and partial sky. In both cases of sky coverage and for both simulated and observed foreground-cleaned CMB maps at $n_{h}=128$, the multipole range under consideration is $[2,256]$. This is because the monopole (which corresponds to the uniform temperature of the CMB) and the dipole (containing contributions from the Doppler shift due to solar motion) must be disregarded for cosmological inferences. For the observed foreground-cleaned full sky CMB map ($m_i$) at resolution $n_h$ in harmonic space, we set the spherical harmonic coefficients corresponding to the monopole and dipole to zero. With the new set of spherical harmonic coefficients, we generate the corresponding full sky CMB map ($m_f$) which is devoid of the monopole and dipole.

Further, since the simulated maps are devoid of any beam smoothing effects, therefore any existing beam smoothing in all the observed foreground-cleaned CMB maps is removed as well. In order to deconvolve beam effects from the observed foreground-cleaned CMB map, we consider the map $m_i$ in harmonic space. We divide all the spherical harmonic coefficients ($a_{(\ell m)i}$'s) of the map by the beam window function ($B_{(\ell)i}$) of the respective observational instrument. With these new spherical harmonic coefficients ($a_{(\ell m)f}$) we construct the full sky map $m_f$ which is devoid of beam smoothing effects. Hence,
\begin{eqnarray}
a_{(\ell m)f} &=& a_{(\ell m)i}\times \frac{B_{(\ell)f}\times P_{(\ell)f}}{B_{(\ell)i}\times P_{(\ell)i}}.
\end{eqnarray}
Here, the initial beam window function $B_{(\ell)i}$ corresponds to a full-width at half-maximum (FWHM) $=5'$ for Planck maps, or $=60'$ for WMAP maps. The final beam window function $B_{(\ell)f}$ corresponds to an FWHM $=0'$ for both Planck and WMAP maps. The initial pixel window function $P_{(\ell)i}$ corresponds to an $n_{side}=2048$ for Planck maps or $n_{side}=512$ for WMAP maps. The final pixel window function $P_{(\ell)f}$ corresponds to an $n_{side}=128$ for both Planck and WMAP maps.

Firstly we consider the case of full sky CMB maps, for which we directly use map $m_f$ to compute the NLVs inside the discs of radius $r_h$ and assign them to pixels of a map at resolution $n_l$. In this case, an ANN is modelled to be trained with $12\times 16^2=3072$ input features. The observed foreground-cleaned CMB maps tested in this case are all the available inpainted ones from \href{https://pla.esac.esa.int/\#maps}{Planck} 2013 \citep{2014A&A...571A..12P} and 2018 \citep{2018PlanckDiffuse} data releases. The application of the ANN to NLV maps obtained from these inpainted maps makes it unlikely to attribute the findings to any minor residual foreground contamination from the galactic region.

Secondly, we consider partial sky CMB maps, obtained after masking with the \href{http://pla.esac.esa.int/pla/aio/product-action?MAP.MAP_ID=COM_Mask_CMB-union_2048_R1.10.fits}{$U73$} mask from Planck 2013 release \citep{2014A&A...571A..12P}, which sufficiently excludes the galactic region in addition to extragalactic point sources. The use of a mask helps minimize contributions from any minor foreground residuals. This $U73$ mask at $n_h=128$ is obtained after downgrading the original binary mask and setting all pixels with values $\geq 0.8$ to one and the rest to zero. Thus for the case of partial sky coverage, we take a map $m_f$ and apply the $U73$ mask, for example in Figure \ref{masked_map}. We calculate the NLVs only for discs which are not masked beyond $90\%$ of their area, following the strategy of \cite{Akrami_2014}. We then assign these NLVs to the map at resolution $n_l$. Thus for partial sky coverage, the ANN architecture is designed to work with $2652$ input features, which are the remaining pixels in the $n_{l}=16$ map, after obeying this criterion for disc rejection. We test the ANN on NLV maps obtained from the observed foreground-cleaned partial sky CMB maps from all releases of \href{https://pla.esac.esa.int/\#maps}{Planck} (2013-2021) \citep{2014A&A...571A..12P, 2015PlanckDiffuse, 2020A&A...641A...6P, 2021Planck}, and \href{https://lambda.gsfc.nasa.gov/product/wmap/current/index.html}{WMAP} (1yr-9yr) \citep{Bennett_2003_1yr, Hinshaw_2007, Hinshaw_2009, Gold_2011, Bennett_2013}.

\section{Training the neural networks}\label{train_process}

\begin{figure*}
    \centering
    \includegraphics[width=0.9\textwidth]{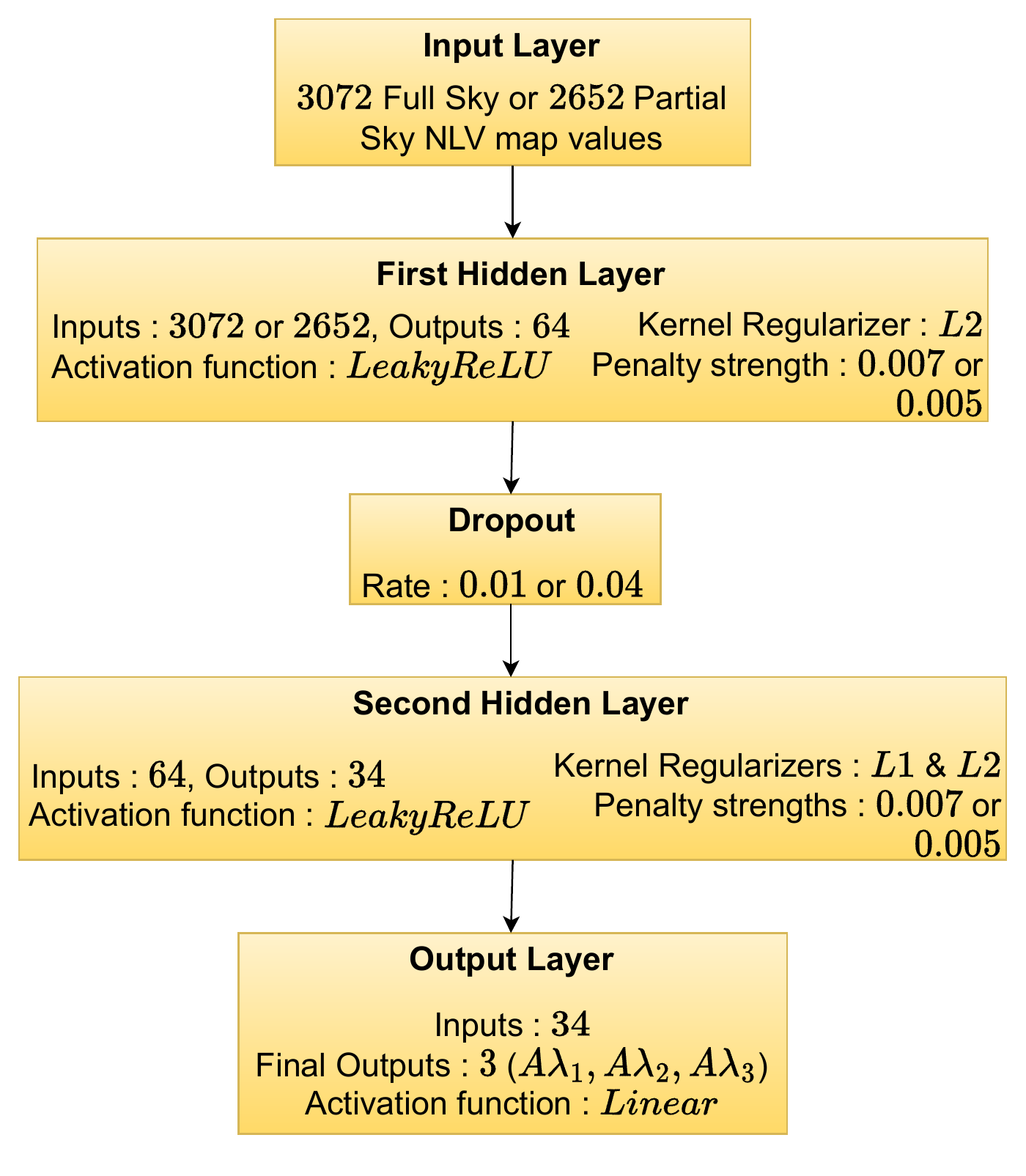}
    \caption{A schematic diagram of the common ANN architecture for detecting the dipolar modulation signal. The differences between full and partial sky cases are mentioned with `or'. All layers after the input layer are densely connected to their preceding layers. The dropout implemented after the first hidden layer has a rate of $0.01$ for full sky or $0.04$ for partial sky. Additionally, the kernel regularizers used are $L2$ in the first hidden layer and both $L1$ and $L2$ in the second hidden layer with strengths of $0.007$ each for the full sky case and $0.005$ each for the partial sky case.}
    \label{common_ann}
\end{figure*}

We have two ANNs, one for each of the full and partial sky cases. The input features used for training the ANNs are values of the NLV map arrays. Both the ANNs are similar in structure, save the difference in the number of input features and hyper-parameters associated with regularization methods. The ANNs are trained on realisations with the same seed values, but with different sky coverage. 

Since $\hat{\lambda}$ is a unit vector, the degrees of freedom in ascertaining the components of $\hat{\lambda}$ are only two. Another degree of freedom in constructing the dipole modulated map is that of $A$. Thus, there are three degrees of freedom in total. Therefore we utilise the three components of the vector $A\times\hat{\lambda}$ as the associated training labels. For SI obeying CMB maps, these three labels are always zero, whereas for SI violating ones, they are non-zero. We choose the training labels in this manner since any other choices of training labels such as those of $(A,\theta, \phi)$ or $(A, \lambda_1, \lambda_2)$ and the like cannot be unambiguously defined for SI obeying maps. 

For the training set from simulated data, we compute the mean and standard deviation of the input features (denoted by $\mu_{in}$, $\sigma_{in}$) as well as those of the training labels (denoted by $\mu_{out}$, $\sigma_{out}$). We re-scale both the inputs and labels by subtracting their respective means from the entire set and dividing the resultant by their respective standard deviations. For similarly re-scaling the validation and test sets for simulated data, we use the previously computed means ($\mu_{in}$, $\mu_{out}$) and standard deviations ($\sigma_{in}$, $\sigma_{out}$) from the training set. Further this scaling is appropriately taken care of for the test set from observed foreground-cleaned CMB data. 

A schematic flowchart to describe the ANN architecture common to both full and partial sky cases is shown in Figure \ref{common_ann}. The differences between the two cases in the input layer and regularization parameters such as the rates of dropout and strengths of penalty for kernel regularizers are mentioned accordingly. We describe the two cases in further detail as follows. 

\subsection{Full Sky ANN}

\begin{figure}
    \centering
    \includegraphics[width=\columnwidth, keepaspectratio]{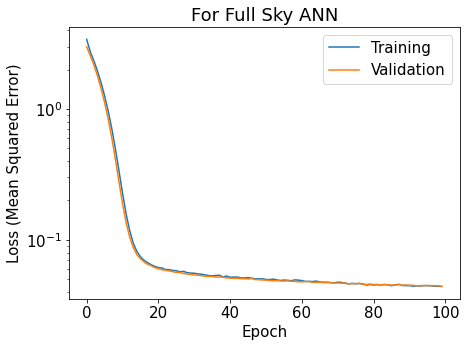}
    \caption{Loss curves for the ANN modelled using full sky CMB maps. The stabilisation around $80$ epochs and beyond indicates that the ANN is trained.}
    \label{mse_full}
\end{figure}

In the full sky case, we consider $3072$ features in the input layer, which is followed by two hidden layers having $64$ and $34$ nodes each. The first hidden layer has an $L2$ kernel regularizer with strength of penalty $=0.007$. There is a dropout at a rate of $0.01$ after this layer. In the second hidden layer, we have both $L1$ and $L2$ kernel regularizers each with strength of penalty values $=0.007$. The output layer has three nodes corresponding to the three components of $A\times \hat{\lambda}$. The activation function used in each of the hidden layers is $LeakyReLU$ whereas that in the output layer is $linear$.

For training the ANN, we use $mse$ as the loss function, while we use $Adam$ for optimization purposes with a learning rate of $10^{-4}$. We see from Figure \ref{mse_full} that the training is accomplished by the end of approximately $80$ epochs, when training with a batch size of $64$. The time taken for a complete run of 100 epochs is $263.18$ seconds, or $\sim 4.4$ minutes on an ordinary CPU.

\subsection{Partial Sky ANN}

\begin{figure}
    \centering
    \includegraphics[width=\columnwidth, keepaspectratio]{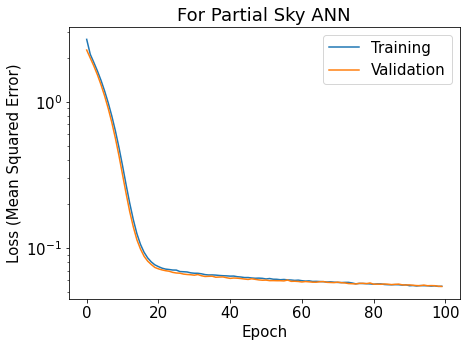}
    \caption{Loss curves for the ANN modelled using partial sky CMB maps. The stabilisation occurs around $80$ epochs and beyond, indicating that the ANN is trained.}
    \label{mse_msk}
\end{figure}

Similar to the full sky case, we have two hidden layers of $64$ and $34$ nodes each. The input layer however takes $2652$ features, which are the remnant pixels on the partial sky NLV map. In this case, the dropout rate used after the first hidden layer is $0.04$. The kernel regularizers have strengths of penalty of $0.005$ for $L2$ at the first hidden layer and $0.005$ for both $L1$ and $L2$ at the second hidden layer. The output layer as usual has three nodes. The activation function is $LeakyReLU$ for both hidden layers, while that for the output layer is $linear$. Again, the $mse$ is used as the loss function, and the optimizer is $Adam$ with a learning rate of $10^{-4}$. The loss curves for training and validation sets are stabilised by $80$ epochs, as seen in Figure \ref{mse_msk}, when the batch size is $64$. For a complete training run of $100$ epochs, the time taken is $254.85$ seconds or $\sim 4.25$ seconds on an ordinary CPU.

\section{Analysis and Results}\label{analysis}

First we present the analysis of test samples with the trained ANNs for the two cases of sky coverage. We specify the goodness of fit for the same with the help of $R^2$ scores for each of the three outputs of $A\lambda_1, A\lambda_2, A\lambda_3$. Mathematically, $R^2$ score can be expressed as
\begin{eqnarray}
    R^2 &=& 1-\frac{\sum \left(y_{true} - y_{pred}\right)^2}{\sum \left(y_{true}-\overline{y_{true}}\right)^2},
\end{eqnarray}where the summations are over all the samples of the set for which the outputs are predicted. It ranges between $0$ and $1$, where $R^2=1$ indicates a perfect fit or the notion that all variations in the predicted data can be explained by the intrinsic dispersion of the actual values. 

Of the total $2\times 10^4$ test samples, half are SI obeying (unmodulated) and the rest are SI violating (modulated) maps. So we can separate them and calculate their respective modulation amplitudes from the predicted outputs as $A=\sqrt{(A\lambda_1)^2+(A\lambda_2)^2+(A\lambda_3)^2}$. We expect the spread of values in amplitude for the unmodulated maps to be very close to zero, and that of modulated maps to closely follow the range $[0.03,0.15]$ in which we have randomly chosen the amplitude.

In the following subsections, we present the respective probability densities of amplitude for unmodulated and modulated maps. Despite our expectation that $A$ from unmodulated maps must be equal to zero, there is a very small non-zero spread in the values of $A$. This is attributable to the fact that the goodness of fit can not be achieved to be exactly equal to one, and is due to the underlying aleatoric uncertainty \citep{10.1007/978-3-030-85469-0_11} of the realisations in question. Hence when we compute $A$ for the observed foreground-cleaned CMB maps, we can say within the confidence defined by this uncertainty, as to whether the predicted value of $A$ from an observed foreground-cleaned CMB map corresponds to a signal of modulation. The significance of detection of the signal is thus quantified with $p$-values of predicted $A$ for observed foreground-cleaned maps versus the null hypothesis prediction for test samples of unmodulated maps.

\subsection{Full Sky}

\begin{figure*}
    \centering
    \includegraphics[width=\textwidth]{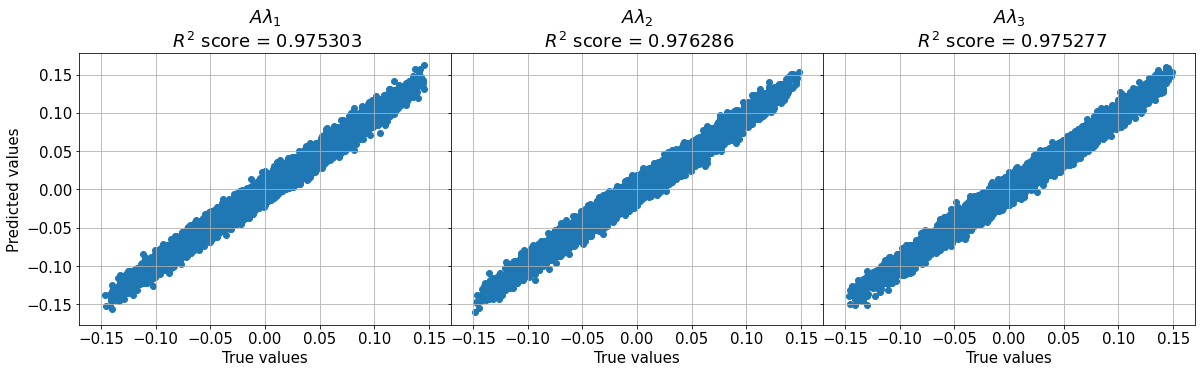}
    \caption{Predicted $A\hat{\lambda}$ components for the test set versus their true values, obtained using the full sky ANN. The predicted values present a good fit to their actual counterparts, as given by $R^2 >\mathbf{0.97}$.}
    \label{test_full}
\end{figure*}
\begin{figure}
    \centering
    \includegraphics[width=\columnwidth]{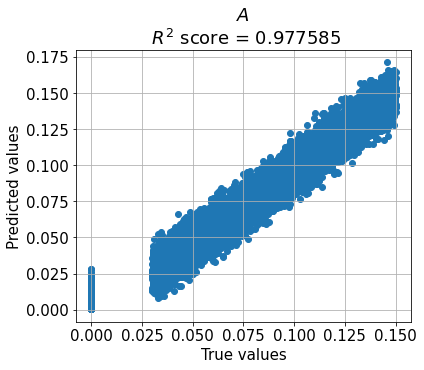}
    \caption{Predicted $A$ for the test set versus their true values, obtained using the full sky ANN. The predicted values present a good fit to their actual counterparts, as given by $R^2 >\mathbf{0.97}$. The amplitude from unmodulated case are isolated on the left of the figure with a dispersion intrinsic to the reconstruction power of the ANN.}
    \label{ampl_full}
\end{figure}
\begin{figure*}
    \centering
    \includegraphics[width=\textwidth]{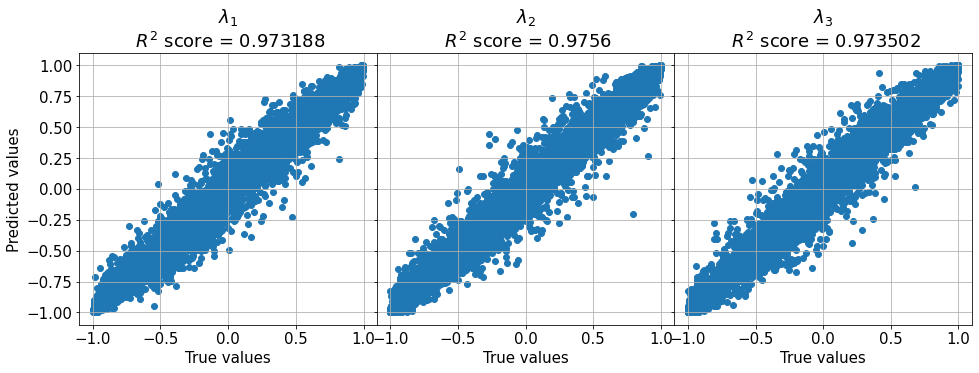}
    \caption{Predicted $\hat{\lambda}$ components for the modulated maps of the test set versus their true values, obtained using the full sky ANN. The predicted values present a good fit to their actual counterparts, as given by $R^2 >\mathbf{0.97}$. The $\hat{\lambda}$ components for unmodulated maps are not shown since they are undefined.}
    \label{lam_full}
\end{figure*}

\begin{figure*}
    \centering
    \includegraphics[width=0.8\textwidth]{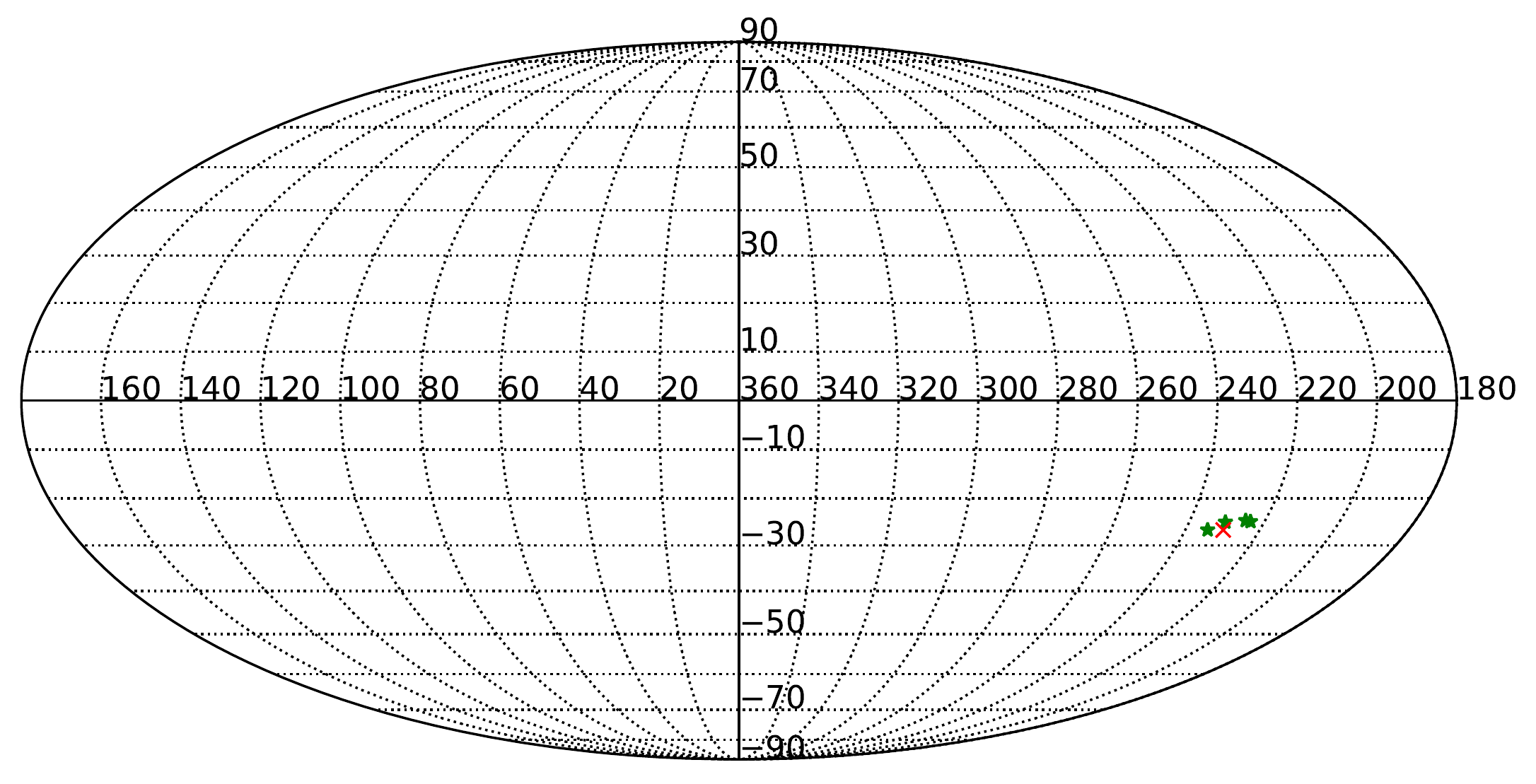}
    \caption{The preferred directions causing a dipolar modulation in the $5$ inpainted observed CMB maps we have investigated with full sky coverage. For the Planck 2013 release inpainted NILC map, the direction is indicated with a red $\times$, while the same for each of the Planck 2018 inpainted CMB maps are shown with a green $\star$. Directions of dipolar modulation in these maps are fairly consistent with each other.}
    \label{map_w_dir_full}
\end{figure*}

\begin{deluxetable*}{ccccc}\label{tab_full}




\tablecaption{Predictions for observed foreground-cleaned CMB maps using the full sky ANN. Both amplitudes and directions for all the maps are similarly valued, and the detection of the dipolar modulation signal is statistically significant.}


\tablehead{\colhead{Map} & \colhead{$A\lambda_1$, $A\lambda_2$, $A\lambda_3$} & \colhead{$A$} & \colhead{Direction ($l$, $b$)} & \colhead{$p$-value}\\ 
\colhead{} & \colhead{} & \colhead{} & \colhead{} & \colhead{}} 

\startdata
COMM 2018 & $ -0.009961 $ , $ -0.012024 $ , $ -0.007281 $ & $ 0.0172 $ & $ 230.3609^\circ $ , $ -25.0006^\circ $  & $ 1.19 \%$  \\
NILC 2013 & $ -0.009646 $ , $ -0.011421 $ , $ -0.007509 $ & $ 0.0167 $ & $ 229.8179^\circ $ , $ -26.6688^\circ $  & $ 1.35 \%$  \\
NILC 2018 & $ -0.012476 $ , $ -0.011948 $ , $ -0.008014 $ & $ 0.019 $ & $ 223.7628^\circ $ , $ -24.8865^\circ $  & $ 0.62 \%$  \\
SMICA 2018 & $ -0.011885 $ , $ -0.01196 $ , $ -0.007744 $ & $ 0.0186 $ & $ 225.1796^\circ $ , $ -24.6688^\circ $  & $ 0.74 \%$  \\
SEVEM 2018 & $ -0.007996 $ , $ -0.011014 $ , $ -0.006827 $ & $ 0.0152 $ & $ 234.0216^\circ $ , $ -26.6376^\circ $  & $ 2.79 \%$  \\
\enddata




\end{deluxetable*}

\begin{figure*}
    \centering
    \includegraphics[width=\textwidth]{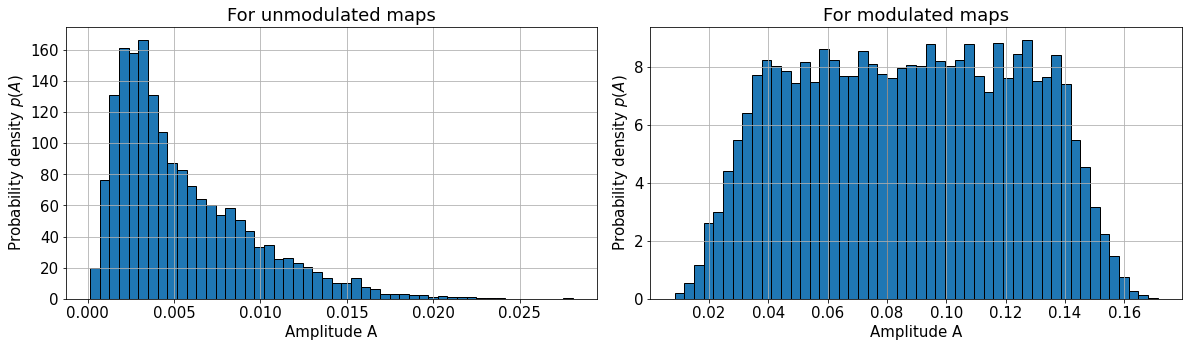}
    \caption{Probability densities of predicted amplitudes $p(A)$ for unmodulated and modulated maps from the test set using the full sky ANN. The histograms closely follow expected ranges of values for unmodulated and modulated maps.}
    \label{amp_full}
\end{figure*}

We test the trained full sky ANN on the $2\times 10^4$ test samples and note the predictions of the ANN for the three components of $A\hat{\lambda}$. These results for the test set are shown along with their goodness of fit scores in Figure \ref{test_full}. The $R^2$ scores are $>\mathbf{0.97}$ indicating that the predicted components of $A\hat{\lambda}$ fit the expected true values quite well. In addition we present the scatter graphs of predicted versus true values of the amplitude ($A$) for the mixed set of unmodulated and modulated maps in Figure \ref{ampl_full}. We show amplitudes for both the unmodulated and modulated maps, and the former can be seen on the left corner of the graph with some dispersion. The scatter graphs of the direction given by the three components of $\hat{\lambda}$ are shown in Figure \ref{lam_full}, only for modulated maps, since they are undefined for unmodulated maps.

The observed foreground-cleaned CMB maps considered in the full sky case are all the available inpainted maps, namely, NILC from Planck 2013 release, and Commander (COMM), NILC, SMICA and SEVEM from Planck 2018 release. We have evaluated the directions for each observed foreground-cleaned CMB map in the following manner. Firstly we normalise the predicted $A\hat{\lambda}$ vector by its respective $A$ to get $\hat{\lambda}$. We then compute $\theta=\cos^{-1}\left(\lambda_3\right)$ and the galactic coordinate $b=90^\circ-\theta$. A general procedure to obtain Galactic $l$ can be outlined as follows:
\begin{enumerate}
\item We must find $\phi=\tan^{-1}\left(\frac{|\lambda_2|}{|\lambda_1|}\right)$.
\item If $\lambda_1 > 0, \lambda_2 > 0$, then $l=\phi$ .
\item If $\lambda_1<0, \lambda_2>0$, then $l=180^\circ-\phi$ .
\item If $\lambda_1<0, \lambda_2<0$, then $l=180^\circ+\phi$ .
\item If $\lambda_1>0, \lambda_2<0$, then $l=360^\circ-\phi$ .
\end{enumerate}The consistency of the preferred directions given by $\hat{\lambda}$ or $(l,b)$ for these five inpainted observed foreground-cleaned CMB maps can be illustrated by plotting the same in a Mollweide map, as shown in Figure \ref{map_w_dir_full}.

Additionally, we present the probability densities of the amplitudes computed from the predicted vector components for the $10^4$ unmodulated and modulated maps of the test set in Figure \ref{amp_full}, which show that the spread in predicted values closely obeys expectations. However, the ANN does not predict a perfect zero for the amplitude in the case of all the $10^4$ unmodulated maps in the test set. Hence, we must gauge the possibility that the ANN predicts a non-zero value for these modulation amplitudes, even if there was no modulation in the observed foreground-cleaned CMB. This is given by a $p$-value which is computed as the percentage of predicted amplitudes from $10^4$ unmodulated maps of the test set that lie above the predicted amplitude for an observed foreground-cleaned CMB map.

In Table \ref{tab_full}, we present the results obtained for these inpainted maps, which lists the direct outputs ($A\hat{\lambda}$ components) from the ANN, the derived values of the amplitude and direction which are consistent across maps, and the $p$-values which indicate a significant detection of the dipolar modulation signal for all the maps.

\subsection{Partial Sky}

\begin{figure*}
    \centering
    \includegraphics[width=\textwidth]{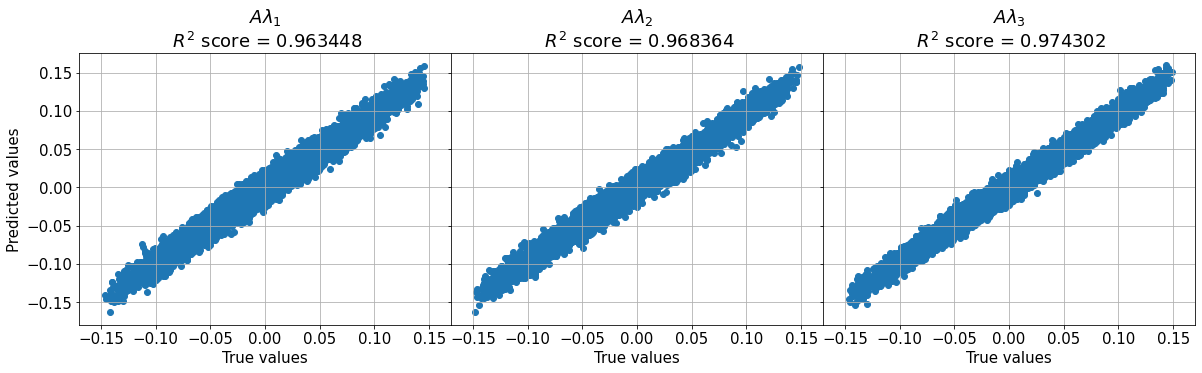}
    \caption{Predicted $A\hat{\lambda}$ components for the test set versus their true values, obtained using the partial sky ANN. The predicted values present a good fit to their actual counterparts, as given by $R^2 >\mathbf{0.96}$.}
    \label{test_msk}
\end{figure*}

\begin{figure}
    \centering
    \includegraphics[width=\columnwidth]{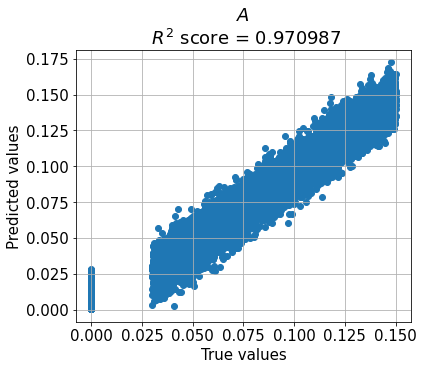}
    \caption{Predicted $A$ for the test set versus their true values, obtained using the partial sky ANN. The predicted values present a good fit to their actual counterparts, as given by $R^2 >\mathbf{0.97}$. The amplitude from unmodulated case are isolated on the left of the figure with a dispersion intrinsic to the reconstruction power of the ANN.}
    \label{ampl_msk}
\end{figure}

\begin{figure*}
    \centering
    \includegraphics[width=\textwidth]{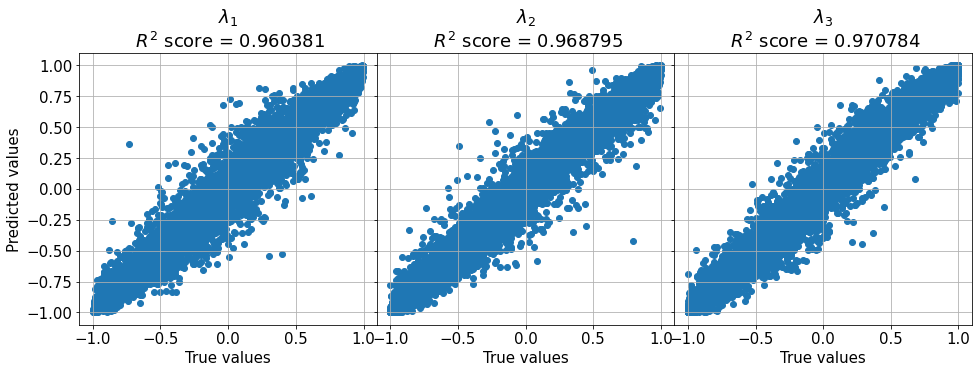}
    \caption{Predicted $\hat{\lambda}$ components for the modulated maps of the test set versus their true values, obtained using the partial sky ANN. The predicted values present a good fit to their actual counterparts, as given by $R^2 >\mathbf{0.96}$. The $\hat{\lambda}$ components for unmodulated maps are not shown since they are undefined.}
    \label{lam_msk}
\end{figure*}

\begin{deluxetable*}{ccccc}\label{tab_msk}




\tablecaption{Predictions for observed foreground-cleaned CMB maps using the partial sky ANN. Overall, both amplitudes and directions across all maps are consistent. Additionally, the detection of the signal of dipolar modulation in all the maps is
statistically significant.}


\tablehead{\colhead{Map} & \colhead{$A\lambda_1$, $A\lambda_2$, $A\lambda_3$} & \colhead{$A$} & \colhead{Direction ($l$, $b$)} & \colhead{$p$-value}\\ 
\colhead{} & \colhead{} & \colhead{} & \colhead{} & \colhead{}} 

\startdata
C-R 2013 & $ -0.022411 $ , $ -0.012969 $ , $ -0.008142 $ & $ 0.0271 $ & $ 210.0573^\circ $ , $ -17.4566^\circ $  & $ 0.04 \%$  \\
COMM 2015 & $ -0.022461 $ , $ -0.013471 $ , $ -0.006697 $ & $ 0.027 $ & $ 210.9529^\circ $ , $ -14.342^\circ $  & $ 0.04 \%$  \\
COMM 2018 & $ -0.022416 $ , $ -0.013744 $ , $ -0.007312 $ & $ 0.0273 $ & $ 211.5149^\circ $ , $ -15.5412^\circ $  & $ 0.04 \%$  \\
NILC 2013 & $ -0.022881 $ , $ -0.014141 $ , $ -0.006798 $ & $ 0.0277 $ & $ 211.7163^\circ $ , $ -14.1835^\circ $  & $ 0.02 \%$  \\
NILC 2015 & $ -0.021521 $ , $ -0.013196 $ , $ -0.007197 $ & $ 0.0263 $ & $ 211.5149^\circ $ , $ -15.9114^\circ $  & $ 0.1 \%$  \\
NILC 2018 & $ -0.022742 $ , $ -0.013616 $ , $ -0.00725 $ & $ 0.0275 $ & $ 210.9099^\circ $ , $ -15.2974^\circ $  & $ 0.02 \%$  \\
SMICA 2013 & $ -0.023005 $ , $ -0.014293 $ , $ -0.006445 $ & $ 0.0278 $ & $ 211.8531^\circ $ , $ -13.3859^\circ $  & $ 0.02 \%$  \\
SMICA 2015 & $ -0.022912 $ , $ -0.013846 $ , $ -0.006701 $ & $ 0.0276 $ & $ 211.1449^\circ $ , $ -14.0539^\circ $  & $ 0.02 \%$  \\
SMICA 2018 & $ -0.022649 $ , $ -0.013983 $ , $ -0.006931 $ & $ 0.0275 $ & $ 211.6895^\circ $ , $ -14.5946^\circ $  & $ 0.02 \%$  \\
SEVEM 2013 & $ -0.022908 $ , $ -0.014313 $ , $ -0.00661 $ & $ 0.0278 $ & $ 211.9983^\circ $ , $ -13.7514^\circ $  & $ 0.02 \%$  \\
SEVEM 2015 & $ -0.021601 $ , $ -0.013769 $ , $ -0.007109 $ & $ 0.0266 $ & $ 212.5148^\circ $ , $ -15.5113^\circ $  & $ 0.07 \%$  \\
SEVEM 2018 & $ -0.021764 $ , $ -0.013886 $ , $ -0.007099 $ & $ 0.0268 $ & $ 212.54^\circ $ , $ -15.376^\circ $  & $ 0.06 \%$  \\
SEVEM 2021 & $ -0.02145 $ , $ -0.013833 $ , $ -0.007017 $ & $ 0.0265 $ & $ 212.8182^\circ $ , $ -15.371^\circ $  & $ 0.07 \%$  \\
WMAP 1yr & $ -0.039227 $ , $ -0.019561 $ , $ -0.00865 $ & $ 0.0447 $ & $ 206.5035^\circ $ , $ -11.1626^\circ $  & $ 0.0 \%$  \\
WMAP 3yr & $ -0.027934 $ , $ -0.015285 $ , $ -0.009217 $ & $ 0.0331 $ & $ 208.6861^\circ $ , $ -16.1434^\circ $  & $ 0.0 \%$  \\
WMAP 5yr & $ -0.022598 $ , $ -0.012087 $ , $ -0.00881 $ & $ 0.0271 $ & $ 208.1404^\circ $ , $ -18.971^\circ $  & $ 0.04 \%$  \\
WMAP 7yr & $ -0.020355 $ , $ -0.012319 $ , $ -0.008 $ & $ 0.0251 $ & $ 211.1818^\circ $ , $ -18.5848^\circ $  & $ 0.17 \%$  \\
WMAP 9yr & $ -0.014497 $ , $ -0.00986 $ , $ -0.007313 $ & $ 0.019 $ & $ 214.2221^\circ $ , $ -22.6406^\circ $  & $ 1.66 \%$  \\
\enddata




\end{deluxetable*}

\begin{figure*}
    \centering
    \includegraphics[width=0.8\textwidth]{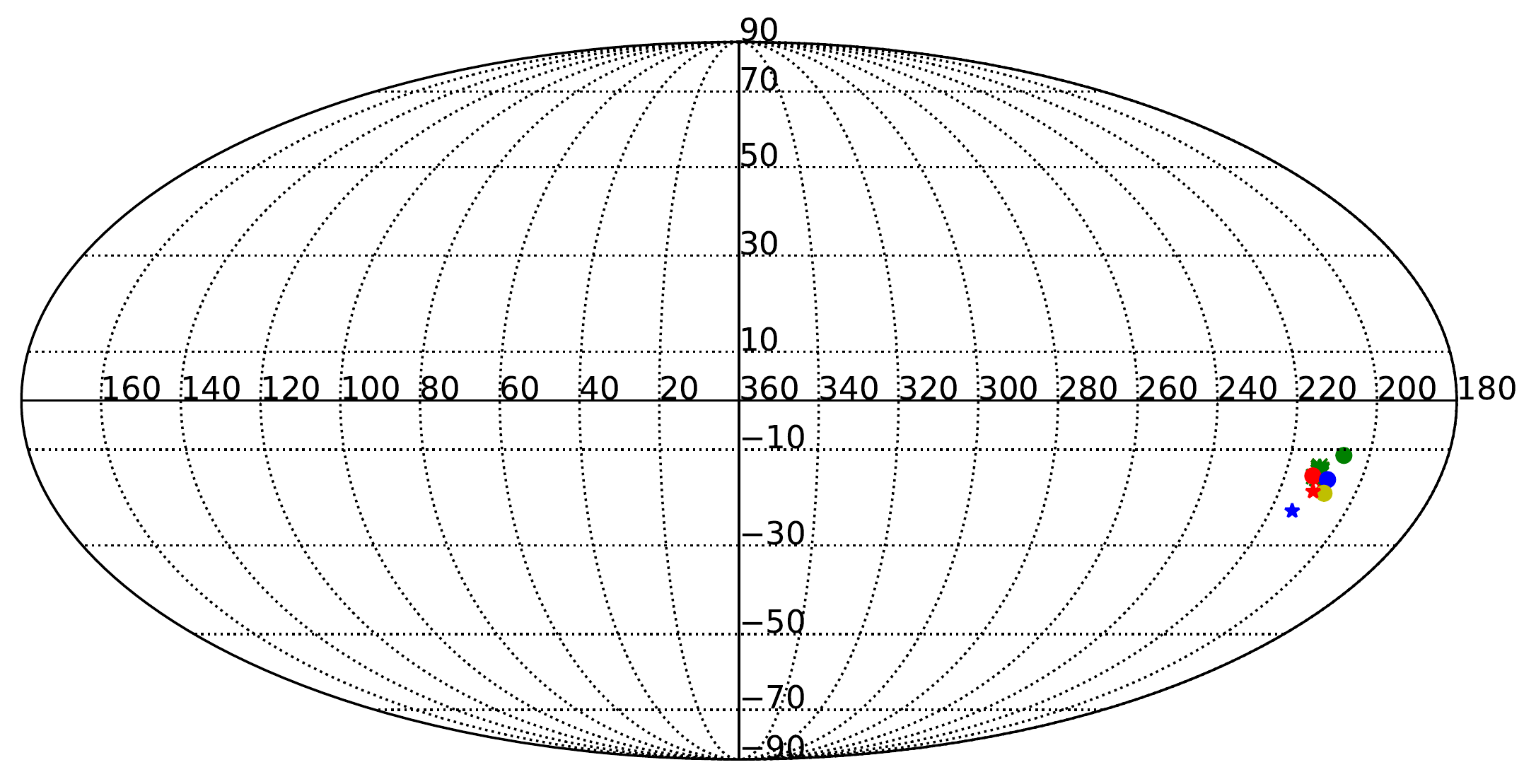}
    \caption{The preferred directions causing a dipolar modulation in the $18$ observed foreground-cleaned CMB maps we have investigated with partial sky coverage. Preferred dipole direction in observed foreground-cleaned partial sky CMB maps from Planck releases of 2013, 2015, 2018, and 2021 are indicated with red $\times$'s, green $\times$'s, green $\star$'s and a red $\bullet$, respectively. Those for each of WMAP 1yr, 3yr, 5yr maps are shown with a green, blue and yellow $\bullet$, and directions for WMAP 7yr and 9yr are shown with a red and a blue $\star$. The figure shows that the directions are mostly consistent over all these variously procured maps.}
    \label{map_w_dir_msk}
\end{figure*}
\begin{figure*}
    \centering
    \includegraphics[width=\textwidth]{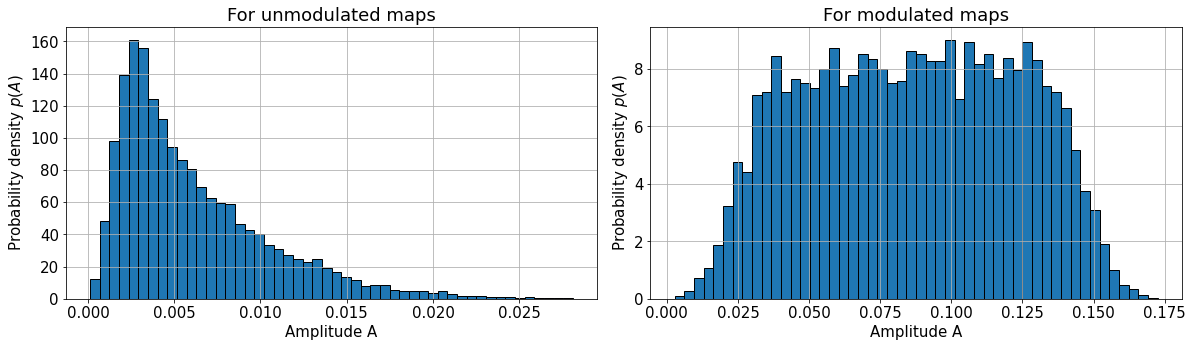}
    \caption{Probability densities of predicted amplitudes $p(A)$ for unmodulated and modulated maps from the test set for the partial sky ANN. The range of predicted amplitudes appropriately follow zero and non-zero values for unmodulated and modulated maps.}
    \label{amp_msk}
\end{figure*}

For the case of partial sky coverage, we apply the corresponding trained ANN on the test set and then on available foreground cleaned CMB maps from all the releases of Planck and WMAP satellites.

In Figure \ref{test_msk}, we present the scatter graph of predicted values of the components of the $A\hat{\lambda}$ vector with respect to the true values, along with their respective $R^2$ scores. Despite the fact that the goodness of fit scores are $>\mathbf{0.96}$, we notice that they are lower than those in the full sky case. This is due to the increased variations that cannot be explained by a similar dispersion in the true values, obviously caused by the use of a mask in this case. Further, since the $U73$ mask primarily conceals galactic sources, it is somewhat symmetric about the $z$-axis, and hence the $R^2$ score for $A\lambda_3$ is not as compromised as those of $A\lambda_1$ and $A\lambda_2$. This is in contrast with the full sky case, for which the goodness of fit values of all the three components were similar (about $\mathbf{0.975}$).

Further we show the scatter graphs of the amplitude ($A$) for the test set in Figure \ref{ampl_msk}. We present amplitudes for both the unmodulated and modulated maps, and the former shows some dispersion for the unmodulated case on the left corner of the figure. Similar to the case of full sky coverage, the three components of $\hat{\lambda}$ are undefined for unmodulated maps. Hence, the scatter graphs for the direction are shown in Figure \ref{lam_msk} only for modulated maps.

When the partial sky ANN is applied to NLV maps of the partial sky observed foreground-cleaned CMB, we see a very consistent amplitude and direction of dipolar modulation across all the maps from Planck and WMAP releases ranging from Planck 2013-2021 data and WMAP 1yr-9yr data. The consistency of the preferred directions given by $\hat{\lambda}$ or $(l,b)$ for the $18$ observed foreground-cleaned CMB maps can be inferred from a plot of the same in a Mollweide map, as presented in Figure \ref{map_w_dir_msk}.

Following our approach in the full sky case, we estimate the amplitudes from predicted $A\hat{\lambda}$'s and present their probability densities for the $10^4$ unmodulated and modulated maps from the test set in Figure \ref{amp_msk}. On finding the minimum and maximum values of predicted amplitudes for these two types of maps, we see that there is a very subtle increase in their spreads (of the orders of $10^{-4}$ for unmodulated maps, and $10^{-3}$ for modulated maps), compared to the full sky case. Nonetheless, these histograms are qualitatively similar to those obtained for the case of full sky coverage, and the predicted amplitudes for unmodulated maps are again not exactly zero. Thus we can quantify the significance of detection for the dipolar modulation signal in the observed foreground-cleaned partial sky CMB maps. We represent this significance with the $p$-value which is computed as the percentage of $10^4$ unmodulated maps for which the predicted amplitudes are larger than those from the observed foreground-cleaned CMB maps. 

We finally present all the findings from the partial sky analysis in Table \ref{tab_msk}, which shows the partial sky ANN outputs for the three $A\hat{\lambda}$ components, the computed values of the amplitudes and directions which are consistent across maps, and the $p$-values which correspond to a significant detection of the signal of dipolar modulation for all the maps.

\section{Summary and conclusion}\label{conc}

The CMB temperature fluctuations are expected to obey statistical isotropy (SI) according to the Standard ($\Lambda CDM$) model of Cosmology. This entails that there must be no preference of a direction in the CMB. However, the hemispherical power asymmetry as seen by many authors in existing literature indicates a departure from the Standard Model. This departure is significant given the reported magnitudes of the $p$-values, and the sheer volume of such findings obtained with independent methods. An underlying dipolar modulation is suggested as a possible cause of this power asymmetry, the strength of which is known to vary with the scale at which it is estimated.

For the first time, we use deep learning with Artificial Neural Networks (ANNs) to probe the existence of a possible dipolar modulation signal. This provides a novel approach towards validating or rejecting evidence for such a signal in previous literature. Employing ANNs for studying features in the CMB may introduce a paradigm shift in interpretation of signals of SI violation, relative to traditional methods of regression or fitting associated with the frequentist approach. This is because ANN architectures can `learn' how to detect a signal when presented with a set of samples for training. Upon adequate training the ANN develops the artificial intelligence to act on observed foreground-cleaned CMB data and consequently estimate a possible signal in such data.

We consider normalised local variance (NLV) maps which are very useful as input features to train ANNs. We build two ANN architectures namely, for the full and partial sky cases. To obtain partial sky maps, we use the Planck $U73$ mask released in $2013$. The key findings of this work are as follows.

\begin{enumerate}
\item With full sky coverage,
\begin{enumerate}
    \item generally consistent values of amplitude and direction of the modulation are seen across all available observed foreground-cleaned full sky inpainted maps from all releases of Planck (COMM 2018, NILC 2013, NILC 2018, SMICA 2018, SEVEM 2018).
    \item The detected signal is significant (at $97.21\%-99.38\%$ C.L.) for all these $5$ maps.
\end{enumerate}
\item With partial sky coverage,
\begin{enumerate}
    \item we find reasonably consistent values of amplitude and direction of the modulation across all observed foreground-cleaned partial sky maps from all releases of Planck (2013-2021) and WMAP (1yr-9yr).
    \item The detected signal is significant at $99.9\%-99.98\%$ C.L. for all the $13$ Planck maps, and at $98.34\%-100.0\%$ C.L. for all the $5$ WMAP maps. 
\end{enumerate}
\item These results are therefore robust against sky coverage, observational instruments, periods of observation, and foreground cleaning and inpainting methods.
\end{enumerate}

In the following paragraphs, we discuss two criticisms that have been posited against the manner in which any possibly anisotropic signals in the CMB are probed, and address how our method is able to mitigate those effects further.

Firstly the look-elsewhere effect occurs when a signal is detected purely by chance and is attributable to the large sample size for which it becomes more favourable to see some random fluctuations that are statistically significant. It is additionally the result of a constant approach of `looking elsewhere' to find a significant signal, while disregarding any previously insignificant findings. In this work, we are able to weaken the look-elsewhere effect (a) with the robustness of the detection, and (b) by adding to existing literature an independent method like the one in this work, which also detects a significant signal, thus strengthening the repeatability of the initial findings.

Secondly, the concept of a posteriori statistical inference is based on devising estimators to shift our focus to visually anomalous features. However, (a) since the method of deep learning to distinguish unmodulated maps from modulated ones (quantified with the magnitude of the modulation) is distinct from the process of devising an estimator which focuses on a search for such a signal after looking at the data, and (b) as we use a wide range of amplitudes and directions to train the ANN so that it is not focused at detection of amplitude and directions specific to the observed foreground-cleaned data and can be used to probe unseen data, we are able to alleviate the criticism of an a posteriori choice of statistics.

In conclusion, we can say that our findings agree quite closely with those in existing literature. Further, assuming that no unknown residual systematics are commonly present in all the observed foreground-cleaned CMB maps considered here, this entails that either our universe is a rare realisation of the Standard Model, or that we live in a statistically anisotropic universe which could be a rather common realisation of a different model.

\section*{Acknowledgements}
We acknowledge the use of the publicly available HEALPix \citep{2005ApJ...622..759G} software package (\url{http://healpix.sourceforge.io}). Our analyses are based on observations from \href{http://www.esa.int/Planck}{Planck}, an ESA science mission with instruments and contributions directly funded by ESA Member States, NASA, and Canada. We acknowledge the use of the \href{https://lambda.gsfc.nasa.gov/}{Legacy Archive for Microwave Background Data Analysis (LAMBDA)}, part of the High Energy Astrophysics Science Archive Center (HEASARC). HEASARC/ LAMBDA is a service of the Astrophysics Science Division at the NASA Goddard Space Flight Center. We acknowledge the use of Python deep learning libraries of TENSORFLOW \citep{tensorflow2015-whitepaper} and KERAS \citep{chollet2015keras}. We have used the online cloud service \href{https://colab.research.google.com/}{Google Colaboratory} hosted by Google. We have used the freely available online diagram maker tool accessible as \url{https://www.diagrams.net/} for Figures \ref{neuron1}, \ref{egANN}, and \ref{common_ann}. The dense connections for Figure \ref{egANN} were produced using the freely available online tool called \href{https://alexlenail.me/NN-SVG/}{NN-SVG}. We thank Srikanta Pal and Pallav Chanda for many useful discussions regarding ANNs. \\


\bibliography{output}{}
\bibliographystyle{aasjournal}



\end{document}